\documentstyle[12pt]{article}

\def\beq{\begin{equation}}
\def\eeq{\end{equation}}
\def\bea{\begin{eqnarray}}
\def\eea{\end{eqnarray}}
\def\beas{\begin{eqnarray*}}
\def\eeas{\end{eqnarray*}}
\def\nn{\nonumber}
\begin{document}
\begin{titlepage}
\begin{center}
{\it P.N.Lebedev Institute preprint} \hfill
FIAN/TD/08-91\\
{\it I.E.Tamm Theory Department} \hfill October 1991\\
\vspace{0.1in}{\Large\bf  The Asymptotics of the Correlations Functions\\
in $(1+1)d$ Quantum Field Theory\\
{}From Finite Size Effects in Conformal Theories}\\[.4in]
{\large  A.Mironov}\\
\bigskip {\it Department of Theoretical Physics \\  P.N.Lebedev Physical
Institute \\ Leninsky prospect, 53, Moscow, 117 924}
\footnote{E-mail address: theordep@sci.fian.msk.su}
\\ \smallskip
\bigskip {\large A.Zabrodin}\\
 \bigskip {\it Institute of Chemical Physics\\ Kosygina st., 117334, Moscow}\\
\end{center}

\newpage
\vskip .3in
\centerline{\bf ABSTRACT}
\begin{quotation}

Using the finite-size effects the scaling dimensions and correlation functions
of the main operators in continuous and lattice models of
1d spinless Bose-gas with
pairwise interaction of rather general form are obtained. The long-wave
properties of these systems can be described by the Gaussian model with central
charge $c=1$. The disorder operators of the extended Gaussian model are found
to correspond to some non-local operators in the {\it XXZ} Heisenberg
antiferromagnet. Just the same approach is applicable to fermionic systems.
Scaling dimensions of operators and correlation functions in the systems of
interacting Fermi-particles are obtained. We present a universal treatment for
$1d$ systems of different kinds which is independent of the exact integrability
and gives universal expressions for critical exponents through the
thermodynamic characteristics of the system.

\end{quotation}
\end{titlepage}

\setcounter{page}2
\setcounter{footnote}0

\section{Introduction}

Recently the new effective method to investigate $1d$ quantum systems using
the finite-size effects in conformal field theory (CFT) was proposed in a
number of
papers [1-6]. In $1d$ quantum systems there exist phase transition at zero
temperature and corresponding long-distance conformal symmetry appears to be
enough to determine the properties independent of the interaction at small
scales. These properties are really of a great interest in the $1d$
systems. Among them the correlation functions, $i.e$. expectation values of
operator products in different points are especially interesting. The
correlations at large distances decrease as a power of the distance (at zero
temperature) the ``critical exponents" being continuously dependent on the
coupling constant [7-9].

In $2d$ systems the conformal invariance is known to constrain the spectrum
of scaling dimensions [10]. It does really determine the critical exponents in
$1d$ quantum systems $(i.e$. in $2d$ models of the quantum field theory
with one space and one time directions). The essential point is that it is
possible to calculate central charge and scaling dimensions of the effective
long-distance conformal theory using the so-called finite-size effects
[1,2,11].

This paper is a review of applications of CFT methods to ``real" physical
$1d$ systems which, as they appear, are neither conformal nor exactly
integrable. We propose a unified treatment of physically relevant $1d$
systems of different kinds. Among them are continuous and lattice models of
Bose or Fermi multiparticle systems with interaction of a quite general
type and $1d$ lattice antiferromagnetic spin chains. We shall study the
long-wave behaviour of the correlation functions in these models. The
corresponding critical exponents are determined from the leading finite-size
corrections to the low energy spectrum. A remarkable point is that these
leading corrections are in a sense universal: they are completely determined
by general thermodynamic characteristics of the systems. So our expressions
for the critical exponents through certain susceptibilities have a wide range
of applicability. They include all previously known examples as particular
cases.

This review is based mostly on our short publications [22,23,43]. Here we
present the more detailed argumentation with new examples and discussions.

To begin with, let us remind the ideology and main formulas of the
finite-size approach in CFT [11]. Consider a CFT on an infinite strip of
width $L$ in
space direction. Then each primary conformal operator $\phi $ gives rise to an
infinite ``tower" of the eigenstates $|\phi _{k,\bar k}
\rangle $ of the Hamiltonian with energies

\beq
E^\phi _L(k,\bar k)=E^{vac}_L+ {2\pi v\over L}(h+k+\bar k)
\eeq
and momenta

\beq
P^\phi _L(k,\bar k)=P^\phi _\infty + {2\pi \over L}(s+k-\bar k).
\eeq
(such that $\langle vac|\phi |\phi _{k,\bar k}
\rangle \neq 0)$. Here $h\equiv \Delta +\bar \Delta $ and
$s\equiv \Delta -\bar \Delta $ are scaling dimension and spin of the operator
$\phi $ respectively, $k,\bar k\geq 0$ - are integers, $E^{vac}_L$ is
the ground state energy of the system in the ``box" of length $L$. The
momentum of the lowest state $|\phi _{0,\bar 0}
\rangle \equiv |\phi \rangle $ of the tower at $L\rightarrow \infty $ is
denoted by $P^\phi _\infty $. At last, the parameter $v$ in (1) takes
into account the difference in units of the space and temporal quantities,
$i.e$. $v$ is merely the velocity of sound excitations (group velocity at the
Fermi surface).

Thus, to find the spectrum of primary operators it is sufficient to calculate
the energies of the lowest excitations
$E^\phi _L(0,0)\equiv E^\phi _L$ of each tower up to the
terms of order $L^{-1}$

\beq
E^\phi _L-E^{vac}_L=2\pi vhL^{-1}.
\eeq
The states from $\phi $-tower with $k,\bar k\neq 0$ correspond to the
so-called descendants (or secondary fields) of $\phi  [10]$. They have the
definite scaling dimensions $h+k+\bar k$ (but they are not primary fields!) and
give a contribution to correlators too. However, we shall see that the leading
term of asymptotics of the correlators is determined only by the primary
fields.

The long-wave asymptotics of two-point equal time correlator of the field
$\phi $ has the form

\beq
\langle \phi (x)\phi (0)\rangle \sim \cos (P^\phi _\infty x)x^{-2h},
\eeq
the oscillating factor $\cos (P^\phi _\infty x)$ being determined by a
gap $P^\phi _\infty $ into the momentum spectrum. For simplicity,
throughout this paper we consider only the equal time correlators. Their
asymptotics is completely determined by the scaling dimensions $h$ of the field
$\phi $. The generalization to the space-time correlators is straightforward.
In this case one also should determine the value of spin $s.$

Indeed, the finite-size approach allows one to find out the central charge of
the Virasoro algebra of the conformal symmetry using the equation similar to
eqs.(1)-(3). Namely, the correction $\sim L^{-1}$ to the ground state energy
of the system is proportional to the central charge $c$ [1,2]. For example, for
the theory with periodic boundary conditions one obtains:

\beq
E^{vac}_L= \epsilon _0L - {\pi cv\over 6L}.
\eeq
Here $\epsilon _0$ is the energy of the ground state per unit length in an
infinite volume. The first term in (5) depends on regularization, but
the second does not. In other words, the first (thermodynamic) term is
determined by ultraviolet behaviour of the system; the leading corrections
being given by the infrared properties.

Usually it is convenient to calculate the central charge by means of another
formula which is equivalent to (5). Due to the modular invariance the
case of zero temperature $(T=0)$ and large $L$ is equivalent to the case
of the infinite volume $(L=\infty )$ and low temperature $T$. In the latter
case (5) is rewritten as follows:

\beq
f (T) = \epsilon _0 - {\pi cT^2\over 6v},
\eeq
where $f (T)$ is the density of the free energy as a function of the
temperature (at small $T).$

In the paper [3-6] the equations (1)-(3),(5)-(6) were used to calculate the
critical exponents in $1d$ exactly solvable models. Namely, the low-lying
excitation energies have been calculated within the Bethe-ansatz framework
[12]. At present time there exist many papers devoted to this topic (see, for
example, [1-6,13-20] and references therein) and we would not repeat here the
finite-size calculations in integrable theories. Instead, we shall concentrate
on the following important but less familiar aspects. First, the method turns
out to be applicable to the models with general interactions and leads to
results of the same completeness as in the exact solvable models. Second, the
results which have been obtained earlier are in fact valid only for bosonic
systems. The critical exponents in fermionic systems differ, generally
speaking, from the bosonic ones and can also be obtained by means of the
finite-size method. At last, the finite-size effects  allow one to calculate
the asymptotics of vacuum expectation values of some non-local operators in the
$1d$ systems. All these questions are the subject of the present paper.

We shall consider the following classes of $1d$ systems. The first one
contains the
continuous models of the spinless Fermi- or Bose-gas with interaction of a
general form, the second quantized Hamiltonian being (the mass of particle is
$1/2):$

\beq
\hat H=
\int ^L_0dx\cdot \partial _x\psi ^\ast (x)\partial _x\psi (x)+
{1\over 2}\cdot g\int \int ^L
_0 \  dxdy\cdot \psi ^\ast (x)\psi ^\ast (y)V(x-y)\psi (x)\psi (%
y) \hbox{  .  }
\eeq
Here $L$ is the length of the system, $V(x)$ is some even pairwise (repulsive)
potential of a rather general form, $g>0$ is the coupling constant. The
operators $\psi ^\ast ,\psi $ satisfy usual equal-time (anti)commutation
relations (to indicate the statistics manifestly we sometimes use the notation
$\psi _B$ or $\psi _F)$. The number of particles in the system $N$ is
conserved; in the thermodynamic limit $N\rightarrow \infty $,
$L\rightarrow \infty $ and $\rho =N/L$ is the equilibrium density. Note that
the potential $V(x)=\delta (x)$ admits the exact solution [12].
In the first quantized framework the Hamiltonian
(7) has the form

\beq
\hat H=-\sum ^N_{i=1}{\partial ^2\over \partial x_i^2}
+ g\sum  ^N_{i<j}V(x_i-x_j).
\eeq
Sometimes we use this language for convenience.

The second case under consideration is the lattice version of (7)-(8):

\bea
H = & - & \sum ^L_{x=1}[\psi^{\dag} (x+1)\psi(x) + \psi^{\dag} (x)
\psi(x+1) - 2\psi^{\dag} (x)\psi(x)]\nn \\
& + & g \sum ^L_{x < y} \psi^{\dag} (x)
\psi(x)\ V(x-y)\ \psi^{\dag}(y)\psi(y)
\eea
Here we put the lattice spacing equal to 1, so $x$ and $L$ become dimensionless
quantities now. If necessary, one can easily restore dependence of the lattice
spacing in final answers.

At last, we shall deal with {\it XXZ} Heisenberg antiferromagnet
(or simply the spin chain) with the lattice Hamiltonian:

\beq
\hat H_{XXZ}=
{1\over 2}\sum ^L_{x=1}(\sigma ^1_x\sigma ^1_{x+1} +\sigma ^2_
x\sigma ^2_{x+1}+\cos\gamma \sigma ^3_x\sigma ^3_{x+1})%
\hbox{, }    0\leq \gamma <\pi .
\eeq
Here $L$ is the number of sites of the lattice, $\sigma ^j$ are usual Pauli
matrices, $\gamma $ is the anisotropy parameter. Isotropic ({\it XXZ})
Heisenberg antiferromagnet corresponds to $\gamma =0$. For the {\it XXZ}
antiferromagnet model there exists an exact solution. Namely, all eigenstates
of the Hamiltonian (10) can be constructed explicitly by the Bethe's method
[12]. The simplest eigenstate with all spins looking at the same direction
$(i.e$. full spin is equal to $L/2)$ is the bare (unphysical) vacuum. The
physical vacuum of the antiferromagnet has the minimal possible value of the
full spin (0 or $1/2$ depending on the parity of number of sites $L)$ and
corresponds to the filled bare vacuum, the number of reversed spins being an
analog of the number of particles in the models (7)-(9).

In fact, after the Jordan-Wigner transformation (that is,
$\psi(x) = ( \prod ^{x-1}
_{j=1} \sigma ^3_j) \sigma^-_x$, $\psi^{\dag}(x) = \sigma^+_x (\prod
^{x-1}_{j=1}
\sigma^3_j)$ (10) can be considered as a special case of (9). The simplest
Hamiltonian (10) is chosen to make possible a comparison with the Bethe
ansatz results. We can also consider more general spin 1/2 chains with
the exchange interaction corresponding to arbitrary potentials in (9).

All these
models have a number of common physical properties. The most important property
is the absence of a gap in the low-energy spectrum. In fact, such a general
statement is not quite correct because one correctly can tune $V(x)$ and
$\rho$ to create a gap in the spectrum of the lattice system (9). We are
convinced, however, that the {\it generic case} with spinless $1d$ systems is
just the gapless spectrum with unique sound velocity\footnote{As for the
systems
with internal degrees of freedom, they have in general a number of
branches of gapless excitations with different sound velocities. This more
complicated case is out of the scope of the present paper.
}. Some comments on this point are contained in the Section 2.

Another important property of models under consideration is that
the central charge calculated from (4) in {\it XXZ} antiferromagnet is equal
to 1 [3,14] and just the same holds for the model (7)-(9). So all these
models are described by the universality class of a Gaussian model [21]. It is
not very amusing since the lowest excitations are nothing but free phonons.

Some comments on long-wave approximation are in order. The proper distance
$x_c$ when individual particles have sense is equal to
$x_c\sim L/N=1/\rho $. At large distances $(x\gg x_c)$ there arises an
effective theory describing the phonon excitations, the information about
interaction at small distances being contained in the only parameter $v$. Then
the low-energy excitations corresponding to the linear dispersion law lead to
effective  scaling and conformal invariance (the $O(2)$-invariance is
trivially restored by the choice of units so that $v=1)$. Therefore, the
deviations from the exact conformal theory can only be discovered at small $L$,
so the series in $L^{-1}$ are correct.

Now we briefly describe the content of the paper. In the Section 2 we calculate
the energies of low-lying excitations for the continuous Bose-gas model
and obtain the spectrum of scaling dimensions for this system. The more
complicated case of lattice systems is considered in the Section 3. Here the
universal expressions for the "compactification radius" in the Gaussian model
through thermodynamic parameters of the system will be obtained.
Then, using all these results, we write down
the asymptotical series for correlation functions (Section 4). The most
important examples (pair density correlator and one-particle density matrix)
are discussed in detail. It turns out that a simple generalization of the
finite-size method allows one to find the asymptotics of vacuum expectation
values of some non-local operators as it is demonstrated in Section 5. Here the
main examples are some special non-local operators in the spin chain. At last,
the Section 6 is devoted to the systems of Fermi-particles. We show how the
finite-size method may be used to calculate the exact critical exponents in
this case.

The Appendix is devoted to a non-trivial example of $1d$ multi-particle systems
with long-ranged interaction --- the so-called Sutherland (or, Calogero) model
[28]. First, we calculate the effective central charge from the low-temperature
behaviour of the specific heat. Second, we compare the exact results for
correlation functions in this model with the asymptotical series obtained from
the CFT-approach. This model has never been discussed in the literature from
the conformal point of view. It seems to us that the detailed exposition of
this
example could be useful.

\section{Scaling dimensions in the Bose-gas model}

Let us consider the Bose-gas described by the Hamiltonian (5). To obtain the
scaling dimensions of different operators it is necessary to classify the
low-lying excitations having zero energy at $L=\infty $. We imply periodic
boundary conditions, $i.e$. the particles live on a circle of the length $L.$

The typical excitation spectrum of the $1d$ spectrum is shown schematically
in Fig.1. One can learn that it is gapless with the energy tending to zero when
the momentum is equal to $2\pi \rho m (m$ is integer). These properties hold
for rather large class of both long-range and short-range potentials $V(x)$.
Indeed, the cases of small and large constants $g$ in (7)-(8) have been studied
in
the literature. In [24], where the case of small $g$ in Fermi systems was
studied perturbatively, it was concluded that the correlators have a power-like
behaviour and no gap emerges. The case of long-range $V(x)$ and large $g$
(strong repulsion) was considered in refs.[25,26,27]. In this case, the absence
of the gap becomes quite obvious. Indeed, independently from $V(x)$, the
particles form a regular ``dynamical lattice" (the so-called Wigner crystal)
with sound-type low-energy excitations. From the other hand, there exist
several exactly solvable models which possess a gapless spectrum. These are,
for example, the model with delta-shape potential and that with the potential
$V(x)=x^{-2}$ (the Sutherland model [28]). Evidently, the gap will not
appear when performing small deformations of the Sutherland potential which do
not entail a qualitative rearrangement of the ground state.

We should investigate the spectrum when $L$ is large but finite. Then the
spectrum is quasi-discrete with the energy gap $\sim L^{-1}$ (see Fig.2). We
are interested in the states with zero energy at $L=\infty $. The states
$|\phi \rangle $ corresponding to the primary conformal operators have the
minimal energies in comparison with that of the neighbouring states. So each
gapless branch of the spectrum ``gives rise" to a primary operator. The higher
excitations are the states of the conformal tower with $k>0$ or $\bar k>0$
in (1).

Certainly, we are able to calculate the energy $E^\phi _L$ of the states
$|\phi \rangle $ when the physical states can be constructed explicitly. This
is the case for the exactly solvable models [12], in particular, for the
Bose-gas with $\delta $-shape interaction. But it turns out that in general
case all that information in fact is not necessary and the energy
$E^\phi _L$ of a ``primary" state $|\phi \rangle $ may be obtained using
simple thermodynamic reasoning.

Here we shall demonstrate this for the system (7)-(8). Let us begin with the
excitations which conserve the number of particles. The simplest suitable
excitation is the creation of a phonon. The minimal possible momentum is
$\pm 2\pi /L$ (the dots $A_0,\bar A_0$ in Fig.2). Evidently,
the energy is equal to $2\pi v/L$, and the gap in the momentum spectrum
is absent: $P^\phi _\infty =0$. Then (1,2) imply
$h=\pm s=1$. Let us denote $\phi ^0_\pm $ the corresponding primary
operators. Then

\beq
\langle \phi ^0_+(x)\phi ^0_+(0)\rangle =\langle \phi ^0_-(x%
)\phi ^0_-(0)\rangle  \sim
x^{-2};\ \ \ \ \langle \phi ^0_+(x)\phi ^0_-(0)\rangle =%
\langle \phi ^0_-(x)\phi ^0_+(0)\rangle =0 \hbox{  .  }
\eeq
\bigskip
A more interesting excitation has the momentum
$P^\phi _\infty =2\pi \rho$  ($A_1$ in Fig.2). This excitation is
produced by the ``rotation" of the whole system with minimal angular momentum
$(i.e$. it can be described by the first level of the ``rotator"). In other
words, this state can be produced from the ground state by transition to
another inertial frame of reference. The periodic boundary conditions imply the
quantization of the velocity of this new frame. More generally, one produces
the family of the states $|\phi _{0,m}\rangle $, $m$ is an integer $(A_2$
and so on in Fig.2). The momenta of these states are equal to
$2\pi \rho m$: $P_{0,m}=2\pi \rho m$, and their energies can be obtained by
considering the uniform motion of the whole system, the momentum of each
particle being equal to $2\pi m/L:$

\beq
\delta E_{0,m}= N(2\pi m/L)^2=
{2\pi v\over L}\cdot {2\pi \rho m^2\over v}.
\eeq
Comparing (3) and (12) we find the dimensions of the operators $\phi _{0,m}:$

\beq
h_{0,m}=2\pi \rho v^{-1}m^2\hbox{, }    m\in {\bf Z}\hbox{; } s=0
\hbox{  .  }
\eeq
\bigskip
The above derivation may seem to be not quite correct since it neglects the
quantum nature of the ground state. But more rigorous arguments lead to the
same result. Indeed, let us turn to another inertial frame and use the first
quantized language for convenience. Then the full $(N$-particle) wave function
is transformed as

\beq
\Psi (\{x_i\})\rightarrow \tilde \Psi (\{x_i\})=\exp (iq\sum ^N_
{k=1}x_k)\cdot \Psi (\{x_i\}).
\eeq
This function should be single-valued, so the momentum $q$ is quantized:
$q=2\pi m/L$, $m\in {\bf Z}$. Now acting to $\tilde \Psi $ by the
Schr\"odinger operator (8) one obtains (12).

The same result can be obtained also in the second quantization formalism. In
this case one should work with the Hamiltonian (7) shifted by chemical
potential term

\beq
\hat H \to \hat H - \mu \hat N \hbox{  ,  }
\eeq
where $\hat N = \int ^L_0 dx\psi ^{\dag}(x)\psi (x)$ is the operator of the
number
of particles,

\beq
\mu = \left. \lim_{L \to \infty} {\partial E_L^{vac} \over \partial N}
\right|_{\rho = const}
\eeq
is the chemical potential. The Galilean boost operator is

\beq
\hat K = \int^L_0 \ dx x \psi ^{\dag}(x) \psi (x)
\eeq
and the transition to another inertial frame can be described by

$$
\hat H \to e^{-iq \hat K} H e^{iq \hat K} \hbox{  .  }
$$
We leave the calculation for the reader.

The states $|\phi ^0_\pm \rangle $, $|\phi _{0,m}\rangle $ are excitations
in the sector with the fixed number of particles. Now let us consider the
excitations which change the number of particles. Now one should work with the
modified Hamiltonian (15). In Bose case the adding of $n$ particles results in
the energy shift

\beq
\delta E^B_{n,0}= {1\over 2}n^2(\partial E^{vac}_L/\partial N^2)=
{n^2\over 2L}(\partial ^2\epsilon _0/\partial \rho ^2)=
{2\pi v\over L}\cdot {vn^2\over 8\pi \rho } \hbox{  ,  }
\eeq
where $\epsilon _{^0}$ again denotes the energy density. Here we used the
well-known thermodynamic relation

\beq
v^2=2\rho (\partial ^2\epsilon _0/\partial \rho ^2) \hbox{  .  }
\eeq
Now we can
obtain the dimensions of the new family of primary operators $\phi _{n,0}:$

\beq
h_{n,0}= {vn^2\over 8\pi \rho }\hbox{, }      s=0.
\eeq
Certainly, it is possible to combine the above excitations, $i.e$. to add $n$
particles and then to ``rotate" the system at the $m-th$ level. In this way we
obtain the operators $\phi _{n,m}$ with the dimensions

\beq
h_{n,m}= {n^2\over R^2}
+ {m^2R^2\over 4}\hbox{; }    n,m\in {\bf Z},
\eeq
where we denote

\beq
R^2=8\pi \rho v^{-1}.
\eeq
The dimensions $h_{0,1}$ and $h_{1,0}$ have been obtained first in
[3] (in this paper the case of $V(x)=\delta (x)$ was
considered), where they were represented in terms of integrable systems theory.
One can check that these dimensions coincide with ours, with the former being
expressed through the sound velocity. It seems to us that the
form of the answer given by (21)-(22) is more convenient, as it has sense for
quite arbitrary potentials, and all the information about potential is
contained in the unique parameter $v$ (or $R).$ In the next section we shall
obtain a universal expression for R valid for lattice systems as well.

In exactly solvable models one can calculate the central charge using the
formula (2). Indeed, both the model with $V(x)=\delta (x)$ in
(7)-(8) and {\it XXZ}-spin chain (10) provide $c=1$ [3].
In the Sutherland-Calogero model $(V(x)=x^{-2} - [28])$ the
dependence $f (T)$ at low temperature can be found using the results of [29],
and (6) yields the answer $c=1$ too (see the Appendix). A more general argument
is the
following. Since the scaling dimensions depend on a continuous parameter, the
central charge should be not less than 1 [30]. Then one can always extract an
``irreducible theory" having $c=1$ from the initial theory [31]. On the other
hand, the central charge greater than 1 implies the number of degrees of
freedom greater than 1. However, it is impossible for the systems under
consideration since we have assumed the gapless spectrum with the only branch
of phonon excitations. The result $c=1$ seems to be quite natural as just such
a theory describes the phonon system in the long-wave limit.

Note that the spectrum of scaling dimensions (21) also implies $c=1$. Indeed,
roughly speaking, almost all conformal theories with $c=1$ are equivalent to
the Gaussian models with the spectrum of dimensions like (21) which can be
parameterized by the unique continuous parameter (the whole classification of
those theories would include the Gaussian models with corresponding orbifold
line and three isolated models in moduli space of the conformal theories [24]).

More precisely, the Gaussian model is the $2d$ free massless scalar field
theory with the action (here $z=x+iy$, $\bar z=x-iy):$

\beq
S_R= {1\over 2\pi } \int   d^2z\partial _z\varphi \partial _{\bar z}
\varphi
\eeq
which is evidently $U(1)$-invariant. The field $\varphi (z,\bar z)$ takes
values in the ``circle" of radius $R$, $i.e$. one identifies $\varphi $ and
$\varphi +2\pi R$. The spectrum of this theory is given by (21) [21,32], with
the parameter $R$ playing the role of the compactification radius in string
theory. The operator $\phi ^0_+ (\phi ^0_-)$ can be identified with
the (anti)chiral $U(1)$-current $\partial _z\varphi  (\partial _{\bar z}
\varphi )$, and $\phi _{n,m}$ can be identified with the exponential of the
free field:

\beq
\phi _{n,m}\rightarrow  :\exp \{ip\varphi +i\bar p\bar \varphi \}:\hbox{  ,  }
\eeq
where $:\ldots:$ denotes the proper normal ordering and the dimensions are

\beq
(\Delta ,\bar \Delta )=(p^2/2,\bar p^2/2)\hbox{, }
h=\Delta +\bar \Delta \hbox{, }    s=\Delta -\bar \Delta ,
\eeq

\beq
(p,\bar p)=(nR^{-1}+mR/2,nR^{-1}-mR/2)\hbox{; }
n,m\in {\bf Z} \hbox{  .  }
\eeq
This theory is invariant under the dual transformation $R\rightarrow 2/R$,
and $R=\sqrt{2}$ is the self-dual point corresponding to the isotropic
({\it XXX}) antiferromagnet $(\gamma =0$ in (10)).

\section{Scaling dimensions in lattice models}

As we saw in the previous section,
the continuous models have an {\it exact} symmetry which allows one to
obtain the answer immediately, namely, they are invariant with respect to
Galilean transformations. In the case of lattice models (9)-(10)
the exact Galilean
invariance is broken. Nevertheless, we shall see that it is possible to apply
the finite-size method in this case too [43]. In this section the most general
relations
between critical exponents and thermodinamic parameters of the system will be
obtained.

Consider the system of spinless particles on $1d$ lattice with the
Hamiltonian (9) modified according to (15).
Then
$\rho  = \partial \epsilon _0/\partial \mu $ is the density of particles at
zero temperature. We imply periodic boundary conditions in (9). It will be
convenient to imagine the lattice as being rolled up into a ring.
Again, we shall assume that there is no gap in the energy spectrum of the
system. In
this case our arguments do not depend on the specific form of $V(x)$.

The operators $\phi ^0_\pm$ can be constructed in the same way as in the
continuous case and their dimensions and correlators are given by the same
formulas (11).

Now we consider the analog of ``rotating states'' $|\phi _{0,m} \rangle$.
By definition, $|\phi _{0,m}\rangle $ is the eigenstate of $H$ with momentum
$2\pi \rho m$ and minimal energy. Let us represent it in the form (it is
convenient to work in the second quantization formalism in this case)

\beq
|\phi _{0,m}\rangle  = \sum _{x_j}\exp (ip_m )\sum ^N_{j+1}x_j)
\Psi _m(x_1,\ldots,x_N) \prod ^N_{k=1} \psi ^{\dag} (x_k)|0\rangle \hbox{  ,  }
\eeq
where $p_m = 2\pi mL^{-1}$, $|0\rangle $ is the bare vacuum, $N$ is the
(conserved) number of particles. When $m = 0 |\phi _{0,0}\rangle  =
|vac\rangle $ and $\Psi _0$ is the ground state coordinate wave function. We
have extracted the exponential factor in (27) in order to make $\Psi _m$
close to $\Psi _0$. This is really the case because $p_m \sim  L^{-1}$
(note that in continuous systems $\Psi _m = \Psi _0$ for all $m).$

Acting to $|\phi _{0,m}\rangle  $ by the Hamiltonian $H $ (9) one can see that
$\Psi _m$ should be the ground state coordinate wave function (in the
$N$-particle sector) of the following Hamiltonian $\hat H(p)$ at $p = p_m:$

\bea
H(p) = H & + & (1- \cos p)\sum ^L_{x=1}(\psi ^{\dag} (x+1) \psi (x)
 + \psi ^{\dag} (x) \psi (x+1) \nn \\
& - & i\sin p \sum ^L_{x=1}(\psi ^{\dag} (x+1)
\psi (x) - \psi ^{\dag} (x)
\psi (x+1) \hbox{  .  }
\eea
Evidently, $H(p)$ is produced from   $H = H(0)$ by the following
transformation $\psi
^{\dag} (x) \to
 \exp (ipx)\psi ^{\dag} (x)
$, $\psi (x) \rightarrow  \exp (ipx)\psi (x)$. Due to the gauge invariance it
means that the uniform magnetic field $B = p$ is applied to the system. In
other
words,
$H(p)$ describes the system (9) in the uniform magnetic field orthogonal to our
ring lattice (Fig.3). Note that the magnetic flux through the ring is equal to
$2\pi m$
(because $p$ is quantized: $p = p_{^m}).$

Passing from $H$ to $H(p)$ we can write for the ground state energy shift
$\epsilon _m$ (up
to $L^{-1})$ the following general relation which is evident from the
thermodynamic reasoning:

\beq
\delta E_{0,m} = \epsilon _m = {1\over 2} Lp^2_m
({\partial ^2\epsilon _0\over \partial \ p^2})_{p=0} =
2\pi vL^{-1}\pi \eta v^{-1}m^2 \hbox{  .  }
\eeq
Here $\epsilon _0$ is the minimal eigenvalue of $H(p)$. When there is no
Galilean invariance the current  $j = i\langle \psi ^{\dag} (x+1)
\psi (x) - \psi ^{\dag} (x)
\psi (x+1)\rangle $ depends on momentum $p$ according to a non-linear
relation. So we have introduced the ``current-momentum" susceptibility $\eta $
in (29):

\beq
\eta  = {\partial ^2\epsilon _0\over \partial \ p^2} =
{\partial j\over \partial p} \hbox{  .  }
\eeq
Of course, the same expression for the energy shift (29) may be obtained by
means of the
perturbation theory,
with {\it exact} value of the $L^{-1}$
order term being given by the
first two orders in $p$ (see (28)).

By comparing (3) and (29) we find the dimensions of the corresponding primary
operators:

\beq
h_{0,m} = \pi \eta v^{-1}m^2 \hbox{  .  }
\eeq
Another way to excite the system is to change the number of particles.
Similarly to (18) we have in the
Bose case

\beq
\delta E^B_{n,0} = {1\over 2}
n^2L^{-1}({\partial ^2\epsilon _0\over \partial \ \rho ^2}) =
{1\over 2} n^2\chi ^{-1}L^{-1} \hbox{  ,  }
\eeq
where

\beq
\chi  = {\partial ^2\epsilon _0\over \partial \ \mu ^2} =
{\partial \rho \over \partial \mu }
\eeq
is the usual susceptibility which is in a sense ``dual" to $\eta $.
(Note that in general (19) is no longer true!) Comparing
with (3) we obtain a new family of primary operators with the following
dimensions:

\beq
h_{n,0} = (4\pi v\chi )^{-1}n^2 \hbox{  .  }
\eeq
\bigskip
Now let us note that the two susceptibilities $\eta ,\chi $ are connected with
the sound velocity $v$ by a universal thermodynamic relation. The simplest way
to see this is to consider the wave equation for long waves. Repeating standard
arguments leading to the wave equation and taking into account the relation
between current and momentum one has

\beq
\chi  {\partial ^2u(x,t)\over \partial t^2} = \eta
{\partial ^2u(x,t)\over \partial x^2}
\eeq
(here $u(x,t)$ is the displacement of medium and $x$ should be considered as a
continuous variable). Hence

\beq
v^2 = \eta \chi ^{-1} \hbox{  .  }
\eeq
Therefore, combining the above excitations and making use of (36) we find the
spectrum of dimensions in the form (21), where

\beq
R^2 = 4 \pi v \chi \hbox{  .  }
\eeq
This is the general expression for the compactification radius and (due to
(21)) for the critical exponents. Note that in the case of the exactly
solvable Hubbard model the critical exponents were expressed through the
susceptibility $\chi$ in the paper [44]. Let us consider some particular cases.

If the continuum limit is performed in (9) the exact Galilean invariance is
restored. In this case

\beq
\chi = {2\rho \over v^2}
\eeq
and (37) coincides with (22).

In the case of the $XXZ$ model (10) we have [14] (after re-interpretation of
$\rho$, $\mu$: $\rho$ now is the total spin of the chain, $\mu$ is external
magnetic field):

\beq
v = {\pi \sin \gamma \over \gamma} \hbox{  ,  }
\eeq

\beq
\chi = {\gamma \over 2\pi(\pi - \gamma ) \sin \gamma} \hbox{  .  }
\eeq
Hence

\beq
R^2 = {2\pi \over \pi - \gamma}
\eeq
which coincides with the Bethe ansatz results. The finite size corrections and
scaling dimensions were studied in the framework of Bethe ansatz in papers
[14,45].

\section{Asymptotics of the correlation functions}

In this section we shall obtain the correlation functions of the models
(7)-(10) in
the asymptotic series form. The most important correlators are density-density
correlator $H(x)=\langle \rho (x)\rho (0)\rangle $,
$\rho (x)=\psi ^\ast (x)\psi (x)$ being the density operator, and the
one-particle density matrix $S(x)=\langle \psi ^\ast (x)\psi (0)\rangle $. In
the spin chain the similar objects are
$H(x)=\langle \sigma ^3_x\sigma ^3 _1\rangle $ and
$S(x)=\langle \sigma ^+_x\sigma ^-_1\rangle $, with
$\sigma ^\pm =\sigma ^1\pm i\sigma ^2.$

The operators $\psi (x)$, $\rho (x)$ have no definite conformal dimensions,
since they do not behave properly under conformal transformations.
Nevertheless, they are local operators and, as such, can be represented as
linear combinations of the primary and descendant operators. For simplicity, we
write this sum at the moment rather symbolically without accounting of the
``descendant contributions" (their role will be discussed below). As we shall
see, to obtain the leading asymptotics it is sufficient to take into account
only the primary operator contributions. Then, the general expression for a
local operator $\hat O(x)$ has a form

\beq
\hat O(x)=\sum  _\phi  C_\phi \exp (iP^\phi _\infty x)\phi (x),
\eeq
where the sum runs over primary operators $\phi $ and $C_\phi $ are some
numerical coefficients. A possible gap in the momentum spectrum leads to
appearance of the oscillating factors $\exp (iP^\phi _\infty x)$ in this
formula. To find the correlation functions of the type
$\langle \hat O(x)\hat O(y)\rangle $ it is sufficient to calculate pair
correlators of primary operators $\phi $ by using the well-known technique of
conformal field theory [10,33]. The most convenient way to do this is to
represent primary fields in the exponential form (24) and to average their
products by performing the conventional Gaussian functional integrals with the
action (23).

The coefficients $C_\phi $ are non-zero if
$\langle vac|\hat O(x)|\phi \rangle \neq 0$ in the thermodynamic limit
(here $\langle vac|$ is the physical vacuum; for brevity sometimes we merely
use the brackets to denote the physical vacuum expectation value). For example,
the results of the previous section yield the following series (for Bose
statistics):

\beq
\psi _B(x)=\sum ^\infty _{m=-\infty }c_m\cdot \exp (2\pi i\rho mx%
)\phi _{1,m}(x),
\eeq

\beq
\psi ^\ast _B(x)=\sum ^\infty _{m=-\infty }c^\ast _m\cdot \exp (-%
2\pi im\rho x)\phi _{-1,-m}(x),
\eeq

\beq
\rho (x)= \phi ^0_+(x)+
\phi ^0_-(x)+\sum ^\infty _{m=-\infty }c'_m\cdot \exp (%
2\pi im\rho x)\phi _{0,m}(x),
\eeq
In the last formula $\phi _{0,0}$ denotes the identity operator. Using these
series one can directly obtain asymptotical expansions of the correlation
functions $H(x)$ and $S(x)$ as well as similar multi-point correlators. Note
that the terms
$\langle \phi ^0_+(x)\phi ^0_+(0)\rangle =\langle \phi ^0_-(x%
)\phi ^0_-(0)\rangle  \sim  x^{-2}$ appear in the
$\rho \rho $-correlator, with the corresponding exponent being equal to the
canonical dimension of the operator $\rho $ (independently of the coupling
constant). The asymptotical series for $H(x)$ always contains such term but in
the most interesting examples this term is non-leading.

Now let us discuss the descendant contributions. Generally speaking, the
expression (42) should be supplemented by the descendant operators for each
primary one satisfying the above selection rule. The conformal Ward identities
imply that the descendant operator $\hat O(x)$ in a correlation function can be
substituted by a number of differentiations of the same correlator in which the
primary field $\phi (x)$ stays instead of its descendant $\hat O(x)$. Thus, the
descendant contributions correspond to the terms with exponents exceeding the
leading ones by a positive integer. For example, the first descendant of
$\phi _{0,1}$ gives a contribution to $H(x)$ of the form $x^{- {1 \over 2}R^2
-1} \sin(2\pi \rho x)$. So, accounting for the descendants
carefully, we are write down the whole asymptotical series
$(x\gg \rho ^{-1}):$

\beq
H(x)-\rho ^2= \sum ^\infty _{k=0}A_kx^{-2-k}+
\sum ^\infty _{m=1}x^{-{1\over2}m^2R^2}\left[ \sum ^\infty _{k=0}A_
{m,k}\cos (2\pi m\rho x + {\pi k\over 2})x^{-k}\right] ,
\eeq

\beq
S(x)=
\sum ^\infty _{m=0}x^{-2/R^2-{1\over2}m^2R^2}\left[ \sum ^\infty _
{k=0}B_{m,k}\cos (2\pi m\rho x + {\pi k\over 2})x^{-k}\right] .
\eeq
Here $A,B$ are some numerical coefficients.

The formulas (46),(47) are in a complete agreement with some exact
results. These are the calculation of the density matrix in impenetrable boson
system $(V(x)=\delta (x)$, $g=\infty $ in (7)-(8)) [34,35] and
the exact result for the correlator $H(x)$ in the Sutherland model at the
special value of the coupling constant $g=4$ [28].
These correlators do really have the form of (46),(47) (Appendix). One can also
compare
these formulas with the answers of the papers [26,36], where the asymptotics of
$H(x)$ and $S(x)$ was obtained by a direct calculation in case of the strongly
repulsive particles with long-range interaction. In those papers the
perturbation theory with respect to the small parameter $g^{-1}$, where $g$
is the coupling constant  have been constructed. The result was that the
parameter $R$ should be determined by the same expression (22) provided that
all orders of the perturbation theory are taken into account. However, $R$ is
small parameter when $g$ is large, so only the ``primary" contributions $(i.e$.
the terms with $k=0$ in (46),(47)) were essential
in refs.[26,36]. The rest terms are negligible on the background of slowly
decreasing contributions $\sim x^{-{1\over2}m^2R^2}$. Just the same
reason does not allow one to obtain the term $\sim x^{-2}$ in
(46) by the method of ref.[26]. Meanwhile, there are models $(e.g$.
the antiferromagnet with $\pi /2 < \gamma <\pi )$ where this term becomes
the leading one.

The last check of (46),(47) is the calculation of non-equal-time correlators,
both the charged and neutral ones, by a new exact method proposed in refs.[37].
To compare these results with ours one has to take into account spins of the
operators and to consider the combinations $x\pm vt$ instead of $x$. If we
suppose the symmetry with respect to exchange
$x+vt\Leftarrow \Rightarrow x-vt$, we reproduce the answers of papers [37,38].

Now we would like to discuss the meaning of the series (46),(47) with unknown
coefficients. At the first glance quite arbitrary function can be represented
in such a form. Nevertheless, the expressions (46),(47) contain an important
information.

First, it turns out that in all cases when the correlators can be obtained
using other methods (see above) the coefficients $A,B$ quickly decrease when
$k\rightarrow \infty $, so the series are well convergent. From the other hand,
the above mentioned approaches as well as exact methods usually reproduce just
several leading terms of (46),(47). Thus these expressions seem to be adequate
to the problem. Second, the subseries corresponding to the descendants (the
sums over $k$ in (46),(47)) can be rolled up to smooth analytic function, but
the sum
over primary fields contains the fractional degrees of $x:$

\beq
H(x)- \rho ^2= {f_0(x)\over x^2} +
\sum ^\infty _{m=1}f_m(x)\cdot x^{-m^2R^2/2},
\eeq

\beq
S(x)= \sum ^\infty _{m=0}\xi _m(x)\cdot x^{-2/R^2-m^2R^2/2}.
\eeq
When $R^2$ is irrational, the number of different types of ramification is
infinite $(x^\alpha $ and $x^\beta $ give the same ramification if $\alpha $
and $\beta $ have the same fractional parts). It apparently implies the
essential singularity. On the contrary, if $R^2$ is rational the correlators
(48),(49) contain the finite number of the ramification types. This case
corresponds to the rational conformal theories [39], and demonstrate the
finiteness of the number of fields which are primary with respect to some large
chiral algebra [40].

The question emerges whether the models described by the rational conformal
theories in the long-wave limit have any peculiar properties. Our conjecture is
that the ground state wave function in such models is analytic function of
every variable on the finite covering of the complex plane (may be with
puncture at $\infty )$. Using the results of refs.[28,29] one can easily check
that this is the case for the Sutherland model%
\footnote{We thank M.Olshanetsky and A.Perelomov for the discussion on this
point.}.

Now we would like to discuss briefly some corrections to the expressions
(46),(47) originated from the deviations of the critical point Hamiltonian from
the fixed point one. That is they can differ by the irrelevant operators which
have their conformal dimensions $h_{irrel}$ greater than 2 (and, therefore,
do not effect the critical behaviour). Then they induce contributions to
(46),(47) decreasing with the distance faster than the leading terms being
suppressed by $x^{-h_{irrel}+2}$. Though the structure of series in this
case is more complicated their global structure is the same in main features.
In particular, the notion of rationality (the finiteness of the ramification
types) remains unchanged. Certainly, in special cases the corrections merely
ought not to appears as it is the case for impenetrable Bose-gas.

More interesting point is the possible logarithmic corrections to leading
asymptotics. The reason of this is the appearance of the additional marginal
operators $(i.e$. operators with dimension (1,1)) which might be given rise in
critical Hamiltonian at some special values of coupling constants (and,
therefore, $R)$. As a rule, the logarithmic corrections emerge when the
symmetry of the system increases [32]. For example, in the
$SU(2)$-symmetrical case of isotropic antiferromagnet in zero magnetic
field the leading term of the correlator
$\langle {\bf \sigma} _x {\bf \sigma} _1\rangle $ given by (46) should be
multiplied
by $(lnx)^{1\over2} [41]$%
\footnote{We are indebted to V.E.Korepin who drew our attention to this
paper.}.
In less symmetrical models the multiplicative corrections to leading terms
are absent, but they can appear in sub-leading terms.

To conclude this section two remarks are in order.

One can easily obtain the multi-point correlators using the expressions
(43)-(45), as primary fields can be realized as exponentials of free scalar
field and can be averaged directly with the Gaussian action $(c=1!)$. The final
expression has a type

\beq
\langle \prod  _i \rho (x_i)\rangle  \sim \sum _{\{m_i\}}
c_{m_i}\exp (i2\pi \rho \sum m_i)\prod _{i<j}|x_i-x_j|^{-{1\over2}m_i m_jR^2}
 + \hbox{non-leading terms}
\eeq
and should be compared with the one which can be obtained by the method of
[36]. Then the primary field contributions are the same in both cases
(descendant terms, as above mentioned, can not be taken into account by this
method).

The second remark concerns the coefficients in the sums (43)-(47). Generally
speaking, the arguments of trigonometric functions in (46),(47) should be added
by some constant (or slowly varying) phase (if $c_m\neq c_{-m}$,
$c^\ast _m\neq -c^\ast _{-m}$, $c'_m\neq c'_{-m}$ and so on).
Nevertheless sometimes this phase is absent. For example, this is the case for
(46) since $c'_m=c'_{-m}$ due to the evident symmetry with respect to
the exchange $q\leftarrow \rightarrow -q$ in the wave function $\tilde \Psi $
of the other inertial frame. Here it would be appropriate to remark that the
exponential operators $\phi _{0,m}+\phi _{0,-m}$ from (16) form closed
operator algebra which corresponds to global $O(2)$-invariance instead of
$U(1)$-invariance as it is in (23).

Thus, we have demonstrated that the long wave correlation properties of bosonic
spinless systems can be described by the Gaussian model with proper
``compactification radius" $R$. The parameter $R$ is defined by the general
formula (37) which gives (22) and (41)
in the cases of continuous bose-gas and Heisenberg antiferromagnet
respectively.

\section{Vacuum expectation values of non-local operators}

In the main body of this section we shall consider the Heisenberg spin chain
(10). Let us consider the following non-local operators in this model:

\beq
S(x,y)\equiv \sum_{j=x}^y \sigma ^3_j=\exp \{i\pi q(x,y)\},
\eeq
$q(x,y)$ being the operator of the number of reversed spins at the sites on the
way from $x$ to $y$, and

\beq
T(x,y)=P(x,x+1)P(x+1,x+2)\ldots P(y-1,y)P(y,x),
\eeq
where $P(x,y)\equiv 1/2(1+{\bf \sigma} _x {\bf \sigma} _y)$ is the operator of
permutation of the spins at the sites $x$ and $y$. The operator $T(x,y)$ is
simply the operator of cyclic permutation of the sites along the segment
$[x,y]: x\rightarrow x+1$, $x+1\rightarrow x+2, \ldots, y-1\rightarrow y$,
$y\rightarrow x.$

Our aim in this section will be to find the asymptotics of the following
correlators
$\langle vac|S(x,y)|vac\rangle $, $\langle vac|T(x,y)|vac\rangle $,
$\langle vac|S(x,y)T(x,y)|vac\rangle $ at $|x-y|\gg 1$. Such non-local
correlators are quite interesting objects from the point of view of the quantum
inverse scattering method [42]. Another application of our results is the
calculation of correlation functions in the system of fermions with spin $1/2$
[27] where one should know the asymptotics of the above non-local correlators.
Due to translational invariance these correlators depend only on $|x-y|.$

We shall show that these correlators have a power-law asymptotics. The
corresponding exponents can be calculated using a natural generalization of the
finite-size method. The idea is the following. As it was shown in [21], the
(extended) Gaussian model has a non-local sector in which one should choose the
numbers $n,m$ in (21) to be half-integers. The corresponding operators, $e.g$.
$\phi _{0,{1\over2}}$, are so-called disorder operators. They can be expressed
through the free field $\varphi $ (see (23)) only in a non-local way. Acting at
the ground state of the extended Gaussian model by such an operator, say
$\phi _{0,{1\over2}}$, one obtains a state in the sector with antiperiodic
boundary conditions $(b.c.)$. Therefore, the correlators of even number of
disorder operators have sense in the sector with periodic $b.c$. A couple of
disorder operators at the points $x,y$ in the spin chain looks like a non-local
operator acting in the segment $[x,y]$. Certainly, that is what allows one to
find the asymptotics of the vacuum expectation values of non-local operators
by CFT.

Let us consider an extended Hilbert space containing all states of the spin
chain with different number of sites and different $b.c$. simultaneously. We
introduce the following operators acting in the extended space: the operator
$a^{\dag}_\beta (x) (\beta =\pm 1/2)$ creating a new site of spin $\beta $
between the
sites $x$ and $x+1$ of the initial chain; the operator $b_\beta (x)$
annihilating the site $x$ (of spin $\beta _{^x})$ in the case of
$\beta _{^x}=\beta $ and giving zero if $\beta _{^x}=-\beta $. Thus,
$a_\beta^{\dag}(x)$ acts from the sector of the extended Hilbert space
corresponding to
the chain with $L$ sites to the sector corresponding to $L+1$ sites. Quite
similarly, $b_\beta (x)$ changes the number of sites from $L$ to $L-1$. The
operator $b_{\pm {1\over2}}$ can be identified with non-local operator
$\phi _{\pm {1\over2},0}$ in extended Gaussian model above. We also introduce
the operator

\beq
s(x)\equiv \prod _{j=x}^L \sigma ^3_j,
\eeq
which connects the sectors corresponding to periodic and antiperiodic $b.c.$

Indeed, let us consider the ground state wave function $\Psi (x_1,\ldots,x_M)$
of the antiferromagnet. Here $x_k$ are the coordinates (integer numbers) of
reversed spins. For periodic $b.c$. one has $\Psi (x_1)=\Psi (x_1+L)$
with other $x_k$ being fixed (for brevity we write down explicitly only
one variable). Let us denote $\tilde \Psi =s(x)\Psi $. In order to
find $\tilde \Psi (x_1+L)$ we shall move $x_1$ and keep other variables
fixed. Suppose we have $k$ reversed spins to the right from the site $x$. Then
one can write for $x_1<x:$

\beq
\tilde \Psi (x_1)= s(x)\tilde \Psi (x_1)= (-1)^k\Psi (x_1).
\eeq
For $x_1>x$ one has the following evident equalities:

\beq
\tilde \Psi (x_1+L)= s(x)\Psi (x_1+L)= (-1)^{k+1}\Psi (x_1+L)=
(-1)^{k+1}\Psi (x_1)= -s(x)\Psi (x_1)= -\tilde \Psi (x_1).
\eeq
So $\tilde \Psi $ belongs to the sector with antiperiodic $b.c.$

Now we can write

\beq
T(x,y)=\sum _\beta a^{\dag}_\beta (x)b_\beta (x),
\eeq

\beq
S(x,y)=s(x)s(y).
\eeq
Let us denote by $|L,+\rangle  (|L,-\rangle )$ the ground state of the spin
chain with $L$ sites and periodic (antiperiodic) $b.c$. (in the thermodynamic
limit these states coincide with $|vac\rangle $, but we are interested in
finite size corrections $\sim L^{-1})$. Working with extended Hilbert space
we can consider the ground states in the sectors with another $L$ and
antiperiodic $b.c$. as excited states over $|vac\rangle \equiv |L,+\rangle $.
Roughly speaking, $a^{\dag}_\beta(x)$ creates an excitation with spin $1/2$
(``one half of magnon"). If $L$
is even then only even number of such excitations can exist, and if $L$ is odd
then there necessarily exist odd number of such ``half-magnons". Clearly. in
the latter case the ground state is two-fold degenerate.

One can easily check that

\beq
\langle L-1,+|b_s(x)|L,+1\rangle \neq 0\hbox{, } \langle L+1,+|a^{\dag}_s(x)
|L,+\rangle \neq 0\hbox{, } \langle L,-|s(x)|L,+\rangle \neq 0,
\eeq
in the thermodynamic limit, $i.e$. the states $|L\pm 1,\pm \rangle $ satisfy to
the selection rule of Section 3. This means that in order to find the
asymptotics
at $|x-y|\gg 1$ of $\langle vac|T(x,y)|vac\rangle = 2\langle
a^{\dag}_{+{1 \over 2}}(x)
b_{+{1\over2}}(y)\rangle $ (obviously, $\langle a^{\dag}_{
+{1\over2}}b_{+{1\over2}}(y)\rangle =\langle a^{\dag}
_{-{1\over2}}(x)b_{-{1\over2}}(y)\rangle )$ and
$\langle vac|S(x,y)|vac\rangle =\langle vac|s(x)s(y)|vac\rangle $ we can
use the formula (4) with the scaling dimensions being determined from the
relation (3). It is sufficient to calculate the energy shifts of the states
$|L\pm 1,+\rangle $ and $|L,-\rangle $ up to the first order in $L^{-1}$ and
find their momenta. This can be done using the well-known exact solution
(Bethe-ansatz) [12,14]. Certainly, one should modify the Hamiltonian as
follows: $\hat H_{XXZ}\rightarrow \hat H_{XXZ}-\epsilon _0 L$ where
$\epsilon _0$ is the energy per one site.
We omit these lengthy calculations here.

The results are as follows. The spectrum of scaling dimensions is given by the
formula (21) with half-integer $n,m$. The parameter $R$ is determined according
to (41). The correlation functions look like

\beq
<vac|S(x,y)|vac>=<vac|\exp \{i\pi q(x,y)\}|vac> \sim
\cos [\pi (x-y)/2]\cdot |x-y|^{-\lambda },
\eeq

\beq
<vac|T(x,y)|vac> \sim  |x-y|^{-\mu },
\eeq

\beq
<vac|T(x,y)\exp \{i\pi q(x,y)\}|vac> \sim
\cos [\pi (x-y)/2]\cdot |x-y|^{-\lambda -\mu }.
\eeq
where

\beq
\mu =R^{-2}/2\hbox{, } \lambda =R^2/8.
\eeq
Note that in the case of isotropic ($XXX$) spin chain $(\gamma =0)$ the
operator $s(x)$ has scaling dimension $1/8$ and coincides with the spin field
[32].

The correlators (59)-(61) appear in the calculation of correlation functions
in the multiparticle spin 1/2 systems as auxiliary objects. In particular,
the result (61) was used in [46] for finding the long-wave asymptotics
of the density matrix in the system of spin 1/2 Fermi particles with strong
interaction.

To conclude this section let us make a remark on the corresponding non-local
operators in the Bose-gas model (7)-(8). An analog of $S(x,y)$ is defined by
the
same formula

\beq
S(x,y)=\exp [i\pi q(x,y)],
\eeq
where $q(x,y)$ is now the operator of number of particles in the segment
$[x,y]$. The leading term of $\langle S(x,y)\rangle
(|x-y|\gg \rho ^{-1})$ can be obtained using the above technique:

\beq
<vac|S(x,y)|vac> \sim  \cos (\pi \rho |x-y|)\cdot |x-y|^{-R^2/8}.
\eeq
It is convenient to introduce an analog of the operator $s(x)$ (53):

\beq
s(x)=\exp [i\pi q(x,L)],
\eeq
which will be useful in the next section. The meaning of the operator
$T(x,y)$ in the continuous case is obscure.

\section{Critical exponents in fermionic systems}

So far, dealing with the model (7)-(8), we implied the case of Bose-statistics.
It is known that the critical exponent of the fermionic field correlator
$S_F(x)=\langle \psi ^\ast _F(x)\psi _F(0)\rangle $ differs from that of
the bosonic one $S_B(x)=\langle \psi ^\ast _B(x)\psi _B(0)\rangle$  [26].
In this section we shall obtain the answer in the framework of the finite-size
method.

There are two different approaches to this problem. The first one makes use of
the fact that the operator $s(x)$ (65) produces the Jordan-Wigner
transformation from bosonic field operators to the fermionic ones:

\bea
\psi _F(x)=\psi _B(x)s(x),
\nn \\
\psi _F ^{\ast} (x) = s^{\ast}(x) \psi_B^{\ast}(x).
\eea
Following the line of the previous section we obtain the result:

\beq
S_F(x) \sim  \cos (\pi \rho x)\cdot x^{-2/R^2-R^2/8}.
\eeq
It is also easy to write down the whole asymptotical series for $S_F(x)$ and
see that it is in a good agreement with that of the paper [36]
obtained by completely different methods.

Less formal arguments are the following. Let $\psi (x_1,\ldots,x_N)$ be the
ground
state wave function of the system (7) or (8) (in the first quantization
language). It
should be an eigenfunction of the full momentum $\hat P$ with an eigenvalue
$P$. So we can write

\beq
\exp (i\hat Pa)\cdot \Psi (x_1,\ldots,x_N)= \Psi (x_1+a,\ldots,x_N+a)=
\exp (iPa)\cdot \Psi (x_1,\ldots,x_N),
\eeq
where $a=L/N$. On the other hand, the cyclic permutation
$x_i\rightarrow x_{i+1}$ results in the sign factor:

\beq
\Psi (x_2,x_3,\ldots,x_N,x_1)=(-1)^{N-1}\Psi (x_1,\ldots,x_N).
\eeq
Setting $x_1=x$, $x_2=x+a$, $x_3=x+2a,\ldots,x_N=x+(N-1)a$ and comparing these
formulas%
\footnote{Here we use the fact that the ground state wave function
$\Psi (x,x+a,x+2a,\ldots)$ is non-zero everywhere (as it have no nodes at
all).}
we obtain the following selection rule for the full momentum:

\beq
\exp (iPL/N)=(-1)^{N-1},
\eeq
or, writing $P=2\pi \rho m$, as before,

\beq
(-1)^{2m}=(-1)^{N-1}.
\eeq
Thus, when $N$ is even, the ground state is two-fold degenerate
$(P=\pm \pi \rho )$. In the case of free fermions this fact is obvious.

Clearly, the neutral excitations with zero energy (``rotations" of the system;
see Sect.2) should have momenta $P_0+2\pi m\rho  (m\in {\bf Z})$ where
$P_0$ is the ground state momentum. Thus, the dimensions of $\phi _{0,m}$
are given by the same formula (13). It is easy to see that this is also
the case when we add even number of particles to the ground state. To avoid
misunderstanding it should be stressed that the dimensions in bosonic and
fermionic cases are generally not equal to each other, the point is that the
sound velocity may be different in the two cases.

But the addition of an old number of particles $n$ changes the parity of $N$,
so, according to (71), we have for $n=1:$

\beq
\delta E^F_{1,0}= 2\pi vL^{-1}(R^{-2}+ R^2/16)=
2\pi vL^{-1}h_{1,{1\over2}},
\eeq
where we have formally used the notation from (21).

Combining the different excitations in accordance with (71), we obtain the
general form of scaling dimensions in the fermionic systems. The spectrum is
given by the formula (21) with the following conditions on $n$ and $m:$

\bea
 a) & \hbox{ when } & n \in  2{\bf Z}\hbox{ , } m \in  {\bf Z};\nn \\
b) & \hbox{ when } &  n \in  2{\bf Z}+1\hbox{ , } m \in  {\bf Z}+1/2 ,\nn \\
& i.e. & m=m'+ {1\over 2}\hbox{, } m'\hbox{ is integer. }
\eea

\section{Concluding remarks}

In this paper we have demonstrated how one can find the long-wave asymptotics
of various correlation functions in one-dimensional field-theoretical models
using the finite-size technique in CFT. The obtained results
are quite general. They are valid both for integrable and non-integrable models
with interaction of a general form.

In conclusion, we would like to say a few words about the meaning of our
results. We have shown that the long-wave correlation properties of the wide
class of one-dimensional spinless systems are described by the Gaussian model,
the simplest conformal field theory. The Gaussian model still contains a free
parameter --- ``compactification radius" $R$.

An important point is that the
value of $R$ is determined by the nature of the {\it short-range} interaction
(of the order of the mean distance between the particles $x_c$). So the
short-range properties of the system are included into the effective
long-range CFT through the only one parameter.

One of the main results of
this paper is the expression for $R$ in concrete systems through their
thermodynamic quantities. Clearly, it immediately gives us all the critical
exponents as functions of thermodynamic parameters of the system. At last, for
a given system it is much easier to solve the thermodynamics than to calculate
correlation functions directly.

\bigskip

We are grateful to V.E.Korepin, A.Yu.Morozov and A.A.Ovchinnikov for fruitful
discussions.

\section{Appendix}
\def\theequation{A\arabic{equation}}
\setcounter{equation}{0}

Let us consider as an example the system of the type (8) with

\beq
V(x) = x^{-2}
\eeq
or

\beq
V(x) = {(2\pi /L)^2 \over \sin^2(2\pi x/L)}
\eeq
(in the thermodynamic limit (A1) is equivalent to (A2)). The model (A1) is
called the Calogero model, (A2) --- the Sutherland model [28]. We call its
thermodynamic  limit the Sutherland-Calogero (SC) model.

In order to find the effective central charge we should investigate the
behaviour of the free energy $f(T)$ at low temperature $T \to 0$. The
thermodynamics of SC model was studied in [29].

First, we have the general thermodynamic relation

\beq
f(T) = \mu \rho - P \hbox{ , }
\eeq
where $P$ is the pressure. In [29] it was shown that the thermodynamics of
SC-model is described by the system of equations

\beq
\pi \beta^{1\over 2} \rho = \int^{\infty}_{x_0} \phi^{1 \over 2}(x) e^x
{dx \over (e^x + \alpha )^2} \hbox{ , }
\eeq

\beq
3\pi \beta^{3\over 2} P = 2\int^{\infty}_{x_0} \phi^{3 \over 2}(x) e^x
{dx \over (e^x + \alpha )^2} \hbox{ , }
\eeq
where $\beta=T^{-1}, \alpha = (1+\sqrt{1+2g})/2$ ($g$ is the coupling
constant in (8)),

\beq
\phi (x) \equiv \beta \mu +x+(1-\alpha )\ln(1+e^{-x}) \hbox{ . }
\eeq
For finding power-like corrections in $\beta^{-1}$ at $\beta \to \infty$ we
can put $x_0 = - \infty$ and expand $sqrt{\phi}$ in a power series. Let

\beq
I_n \equiv I_n(\alpha) = \int^{\infty}_{-\infty} dx
\frac{e^x}{(e^x+\alpha )^2}\left[ x+ (1-\alpha )\ln(1+e^{-x})\right]^n\hbox{ ,
}
\eeq
then we rewrite (A4) up to $\beta ^{-2}$ as follows:

\beq
\alpha \pi \rho = mu^{1 \over 2} + \frac{\alpha I_1}{2\beta \mu ^{1 \over 2}} -
\frac{\alpha I_2}{2\beta ^2 \mu ^{3 \over 2}}\hbox{ , }
\eeq

\beq
{3 \over 2}\alpha \pi \rho = mu^{3 \over 2} +
\frac{3\alpha \mu^{1 \over 2}I_1}{2\beta} +
\frac{3\alpha I_2}{8\beta ^2 \mu ^{1 \over 2}}\hbox{ . }
\eeq
The computation of the integrals gives

\beq
I_1 = 0\hbox{ , } I_2 = {\pi^2 \over 3}\hbox{ , }
\eeq
then

\beq
f(T) = \frac{\alpha ^2\pi ^2\rho ^3}{3} - \frac{T^2}{12\alpha \rho } +
o(T^2)\hbox{ . }
\eeq
The sound velocity is [28,29]

\beq
v=2\pi \alpha \rho \hbox{ , }
\eeq
so, comparing with (6), we conclude that $c=1$.

The exact density-density correlation function $H(x)$ is known in this model at
$g=4$ [28]:

\beq
H(x) = F^2(2x) - {1 \over 2}F'(2x) \int^{2x}_0 dz F(z) +1 \hbox{ , }
\eeq
where

\beq
F(x) = {\sin \pi x \over x}
\eeq
(we put $\rho =1$). In this case $\alpha =2$ and $R^2 = 2$.

Expanding (A13) in a power series of $x^{-k}$ we obtain exactly the series (46)
which follows from the CFT.

\bigskip
\bigskip
\begin{center}{{\Large\bf References}}\end{center}
\medskip
1. Bl\"ote H.W., Cardy J.L., Nightingale M.P. Phys.Rev.Lett. 56 (1986) 742\\
2. Affleck I. Phys.Rev.Lett. 56 (1986) 746\\
3. Bogolubov N.M., Isergin A.G., Reshetikhin N.Yu. Pis'ma v ZhETF, 44 (1986)
405; J.Phys.Math.Gen. A20 (1987) 5361\\
4. de Vega H.J., Karowski M. Nucl.Phys. B285 [FS19] (1987) 619\\
5. von Gehlen G., Rittenberg V. J.Phys. A20 (1987) 2577\\
6. Alcaraz F., Barber M., Batchelor M. Ann.Phys. 182 (1988) 280\\
7. Efetov K.B., Larkin A.I. ZhETF, 69 (1975) 764\\
8. Luther A., Peshel I. Phys.Rev. B12 (1975) 3908\\
9. Haldane F.D.M. J.Phys. C14 (1981) 2585\\
10. Belavin A.A., Polyakov A.M., Zamolodchikov A.B. Nucl.Phys. B241 (1984)
333\\
11. Cardy J.L. Nucl.Phys. B270 [FS16] (1986) 186\\
12. See, for example, Thacker H.B. Rev.Mod.Phys. 53 (1981) 253\\
13. Bogoliubov N.M., Izergin A.G., Reshetikhin N.Yu. J.Phys. A20 (1987) 5361\\
14. Alcaraz F., Barber M., Batchelor M. Ann.Phys. 182 (1988) 280\\
15. Alcaraz F., Baake M., Grimm U., Rittenberg V. J.Phys. A21 (1988) L117\\
16. Woynarovich F. Phys.Rev.Lett. 59 (1987) 259\\
17. Berkovich A., Murthy G. Stony Brook preprint, ITP-SB-88-5, 1988\\
18. Alcaraz F., Martins M. J.Phys. A21 (1988) 1\\
19. Johannesson H. J.Phys. A21 (1988) L611\\
20. Izergin A.G., Korepin V.E., Reshetikhin N.Yu. Stony Brook preprint
ITP-SB-88-47,

  1988\\
21. Kadanoff L.P. Ann.Phys. 120 (1979) 39

Kadanoff L.P., Brown A.C. Ann.Phys. 121 (1979) 318\\
22. Mironov A.D., Zabrodin A.V. J.Phys. A23 (1990) L493\\
23. Mironov A.D., Zabrodin A.V. Phys.Rev.Lett. 66 (1991) 534\\
24. Krivnov V.Ya., Ovchinnikov A.A. JETP, 49 (1979) 328\\
25. Krivnov V.Ya., Ovchinnikov A.A. JETP, 55 (1982) 1628\\
26. Ovchinnikov A.A., Zabrodin A.V. JETP, 63 (1986) 1326\\
27. Ovchinnikov A.A., Zabrodin A.V. Phys.Lett. A130 (1989) 139\\
28. Sutherland B. J.Math.Phys. 12 (1971) 246,251

  Calogero F. J.Math.Phys. 10 (1969) 2191\\
29. Krivnov V.Ya., Ovchinnikov A.A. Teor.Math.Phys. 50 (1982) 155\\
30. Friedan D., Qiu Z., Shenker S. Phys.Rev.Lett. 52 (1984) 1575\\
31. Cardy J.L. J.Phys. A20 (1987) L891\\
32. Dijkgraaf R., Verlinde E., Verlinde H. Comm.Math.Phys. 115 (1988) 649\\
33. Dotsenko Vl.S., Fateev V.A. Nucl.Phys. B240 [FS12] (1984) 312\\
34. Vaidya H.G., Tracy C.A. Phys.Rev.Lett. 42 (1979) 3\\
35. Jimbo M., Miwa T., Mori Y., Sato M. Physica, D1 (1980) 80\\
36. Zabrodin A.V., Ovchinnikov A.A. ZhETF, 88 (1985) 1233\\
37. Its A.R., Isergin A.G., Korepin V.E., Slavnov N.A. Australian National
University

preprint, IC/89/107, 120, 139\\
38. Bogoliubov N.M., Izergin A.G., Korepin V.E. Nucl.Phys. B275 [FS17] (1986)
687\\
39. Moore G., Anderson G. Rationality in conformal field theory.
IASSNS-HEP-87/69\\
40. Moore G., Seiberg N. Nucl.Phys. B313 (1989) 16\\
41. I.Affleck, D.Gepner, H.J.Schulz, T.Ziman J.Phys. A22 (1989) 511\\
42. Korepin V.E. Funk.Analiz i priloz. 23 (1989) 15\\
43. Zabrodin A.V. JETP Lett., 51 (1990) 614\\
44. Bogoliubov N.M., Korepin V.E. Int.Journ.Mod.Phys. B3 (1989) 427\\
45. Pokrovsky S.L., Tsvelik A.M., ZheTF, 93 (1987) 2232\\
46. Zabrodin A.V., Ovchinnikov A.A. Teor.Mat.Fiz. 85 (1990) 443
\newpage
\begin{center}{{\Large\bf Figure Captions}}\end{center}

\medskip
{\bf Fig.1} The excitation spectrum of $1d$ spinless Bose-gas (schematically)

\bigskip
{\bf Fig.2} The spectrum of the same system in a finite volume

\bigskip
{\bf Fig.3} The lattice system in the uniform magnetic field
\end{document}